# Experimental Investigation of Temperature-Dependent Gilbert Damping in Permalloy Thin Films

Yuelei Zhao[1,2†], Qi Song[1,2†], See-Hun Yang[3], Tang Su[1,2], Wei Yuan[1,2], Stuart S. P. Parkin[3,4], Jing Shi[5*], and Wei Han[1,2*]

[1]International Center for Quantum Materials, Peking University, Beijing, 100871, P. R. China

[2]Collaborative Innovation Center of Quantum Matter, Beijing 100871, P. R. China

[3]IBM Almaden Research Center, San Jose, California 95120, USA

[4]Max Planck Institute for Microstructure Physics, 06120 Halle (Saale), Germany

[5]Department of Physics and Astronomy, University of California, Riverside, California 92521, USA

†These authors contributed equally to the work

*Correspondence to be addressed to: jing.shi@ucr.edu (J.S.) and weihan@pku.edu.cn (W.H.)

## Abstract

The Gilbert damping of ferromagnetic materials is arguably the most important but least understood phenomenological parameter that dictates real-time magnetization dynamics. Understanding the physical origin of the Gilbert damping is highly relevant to developing future fast switching spintronics devices such as magnetic sensors and magnetic random access memory. Here, we report an experimental study of temperature-dependent Gilbert damping in permalloy (Py) thin films of varying thicknesses by ferromagnetic resonance. From the thickness dependence, two independent contributions to the Gilbert damping are identified, namely bulk damping and surface damping. Of particular interest, bulk damping decreases monotonically as the temperature decreases, while surface damping shows an enhancement peak at the temperature of ~50 K. These results provide an important insight to the physical origin of the Gilbert damping in ultrathin magnetic films.

## Introduction

It is well known that the magnetization dynamics is described by the Landau-Lifshitz-Gilbert equation with a phenomenological parameter called the Gilbert damping (α),[1,2]:

$$\frac{d\vec{M}}{dt} = -\gamma \vec{M} \times \vec{H}_{eff} + \frac{\alpha}{M_S} \vec{M} \times \frac{d\vec{M}}{dt} \tag{1}$$

where $\vec{M}$ is the magnetization vector, $\gamma$ is the gyromagnetic ratio, and $M_S = \left|\vec{M}\right|$ is the saturation magnetization. Despite intense theoretical and experimental efforts[3-15], the microscopic origin of the damping in ferromagnetic (FM) metallic materials is still not well understood. Using FM metals as an example, vanadium doping decreases the Gilbert damping of Fe[3] while many other rare-earth metals doping increases the damping of permalloy (Py)[4-6,16]. Theoretically, several models have been developed to explain some key characteristics. For example, spin-orbit coupling is proposed to be the intrinsic origin for homogenous time-varying magnetization[9]. The *s-d* exchange scattering model assumes that damping results from scattering of the conducting spin polarized electrons with the magnetization[10]. Besides, there is the Fermi surface breathing model taking account of the spin scattering with the lattice defects based on the Fermi golden rule[11,12]. Furthermore, other damping mechanisms include electron-electron scattering, electron-impurity scattering[13] and spin pumping into the adjacent nonmagnetic layers[14], as well as the two magnon scattering model, which refers to that pairs of magnon are scattered by defects, and the ferromagnetic resonance (FMR) mode moves into short wavelength spin waves, leading to a dephasing contribution to the linewidth[15]. In magnetic nanostructures, the magnetization dynamics is dictated by the Gilbert damping of the FM materials which can be simulated by micromagnetics given the boundaries and dimensions of the nanostructures. Therefore, understanding the Gilbert damping in FM materials is particularly important for characterizing and controlling ultrafast responses in magnetic nanostructures that are highly relevant to spintronic applications such as magnetic sensors and magnetic random access memory[17].

In this letter, we report an experimental investigation of the Gilbert damping in Py thin films via variable temperature FMR in a modified multi-functional insert of physical property measurement system with a coplanar waveguide (see methods for details). We choose Py thin films since it is an interesting FM metallic material for spintronics due to its high permeability, nearly zero magnetostriction, low coercivity, and very large anisotropic magnetoresistance. In our study, Py thin films are grown on top of ~25 nm $SiO_2$/Si substrates with a thickness (*d*) range of 3-50 nm by magnetron sputtering (see methods for details). A capping layer of TaN or $Al_2O_3$ is used to prevent oxidation of the Py during measurement. Interestingly, we observe that the Gilbert damping of the thin Py films (*d* <= 10 nm) shows an enhanced peak at ~ 50 K, while thicker films (*d* >= 20 nm) decreases monotonically as the temperature decreases. The distinct low-temperature behavior in the Gilbert damping in different thickness regimes indicates a pronounced surface contribution in the thin limit. In fact, from the linear relationship of the Gilbert damping as a function of the 1/*d*, we identify two contributions, namely bulk damping and surface damping. Interestingly, these two contributions show very different temperature dependent behaviors, in which the bulk damping decreases monotonically as the temperature decreases, while the surface damping indicates an enhancement peak at ~ 50 K. We also notice that the effective magnetization shows an increase at the same temperature of ~50 K for 3 and 5 nm Py films. These observations could be all related to the magnetization reorientation on the Py surface at a certain temperature. Our results are important for theoretical investigation of the physical origins of Gilbert damping and also useful for the purpose of designing fast switching spintronics devices.

## Results and Discussion

Figure 1a shows five representative curves of the forward amplitude of the complex transmission coefficients ($S_{21}$) vs. in plane magnetic field measured on the 30 nm Py film with TaN capping at the frequencies of 4, 6, 8, 10 and 12 GHz and at 300 K after renormalization by subtracting a constant background. These experimental results could be fitted using the Lorentz equation[18]:

$$S_{21} \propto S_0 \frac{(\Delta H)^2}{(\Delta H)^2 + (H - H_{res})^2} \tag{2}$$

where $S_0$ is the constant describing the coefficient for the transmitted microwave power, $H$ is the external magnetic field, $H_{res}$ is the magnetic field under the resonance condition, and $\Delta H$ is the half linewidth. The extracted $\Delta H$ vs. the excitation frequency ($f$) is summarized in Figures 1b and 1c for the temperature of 300 K and 5 K respectively. The Gilbert damping could be obtained from the linearly fitted curves (red lines), based on the following equation:

$$\Delta H = (\frac{2\pi}{\gamma})\alpha f + \Delta H_0 \tag{3}$$

in which $\gamma$ is the geomagnetic ratio and $\Delta H_0$ is related to the inhomogeneous properties of the Py films. The Gilbert damping at 300 K and 5 K is calculated to be 0.0064 ± 0.0001 and 0.0055 ± 0.0001 respectively.

The temperature dependence of the Gilbert damping for 3-50 nm Py films with TaN capping layer is summarized in Figure 2a. As *d* decreases, the Gilbert damping increases, indicative of the increasing importance of the film surfaces. Interestingly, for thicker Py films (e.g. 30 nm), the damping decreases monotonically as the temperature decreases, which is expected for bulk materials due to suppressed scattering at low temperature. As *d* decreases down to 10 nm, an enhanced peak of the damping is observed at the temperature of ~ 50 K. As *d* decreases further, the peak of the damping becomes more pronounced. For the 3 nm Py film, the damping shows a slight decrease first from 0.0126 ± 0.0001 at 300 K to 0.0121 ± 0.0001 at 175 K, and a giant enhancement up to 0.0142 ± 0.0001 at 50 K, and then a sharp decrease back down to 0.0114 ± 0.0003 at 5 K.

The Gilbert damping as a function of the Py thicknesses at each temperature is also studied. Figure 2b shows the thickness dependence of the Py damping at 300 K. As *d* increases, the Gilbert damping decreases, which indicates a surface/interface enhanced damping for thin Py films[19]. To separate the damping due to the bulk and the surface/interface contribution, the damping is plotted as a function of 1/*d*, as shown in Figure 2c, and it follows this equation as suggested by theories[19-21].

$$\alpha = \alpha_B + \alpha_S \left(\frac{1}{d}\right) \quad (4)$$

in which the $\alpha_B$ and $\alpha_S$ represent the bulk and surface damping, respectively. From these linearly fitted curves, we are able to separate the bulk damping term and the surface damping term out. In Figure 2b, the best fitted parameters for $\alpha_B$ and $\alpha_S$ are 0.0055 ± 0.0003 and 0.020 ± 0.002 nm. To be noted, there are two insulating materials adjacent to the Py films in our studies.

This is very different from previous studies on Py/Pt bilayer systems, where the spin pumping into Pt leads to an enhanced magnetic damping in Py. Hence, the enhanced damping in our studies is very unlikely resulting from spin pumping into $SiO_2$ or TaN. To our knowledge, this surface damping could be related to interfacial spin flip scattering at the interface between Py and the insulating layers, which has been included in a generalized spin-pumping theory reported recently[21].

The temperature dependence of the bulk damping and the surface damping are summarized in Figures 3a and 3b. The bulk damping of Py is ~0.0055 at 300 K. As the temperature decreases, it shows a monotonic decrease and is down to ~0.0049 at 5 K. These values are consistent with theoretical first principle calculations[21-23] and the experimental values (0.004-0.008) reported for Py films with $d \geq 30$ nm[24-27]. The temperature dependence of the bulk damping could be attributed to the magnetization relaxation due to the spin-lattice scattering in the Py films, which decreases as the temperature decreases.

Of particular interest, the surface damping shows a completely different characteristic, indicating a totally different mechanism from the bulk damping. A strong enhancement peak is observed at ~ 50 K for the surface damping. Could this enhancement of this surface/interface damping be due to the strong spin-orbit coupling in atomic Ta of TaN capping layer? To investigate this, we measure the damping of the 5 nm and 30 nm Py films with $Al_2O_3$ capping layer, which is expected to exhibit much lower spin-orbit coupling compared to TaN. The temperature dependence of the Py damping is summarized in Figures 4a and 4b. Interestingly, the similar enhancement of the damping at ~ 50 K is observed for 5 nm Py film with either $Al_2O_3$ capping layer or TaN layer, which excludes that the origin of the feature of the enhanced damping at ~50 K results from the strong spin-orbit coupling in TaN layer. These results also indicate that the mechanism of this feature is most likely related to the common properties of Py with TaN and $Al_2O_3$ capping layers, such as the crystalline grain boundary and roughness of the Py films, etc.

One possible mechanism for the observed peak of the damping at ~50 K could be related to a thermally induced spin reorientation transition on the Py surface at that temperature. For example, it has been shown that the spin reorientation of Py in magnetic tunnel junction structure happens due to the competition of different magnetic anisotropies, which could give rise to the peak of the FMR linewidth around the temperature of ~60 K[28]. Furthermore, we measure the effective magnetization ($M_{eff}$) as a function of temperature. $M_{eff}$ is obtained from the resonance frequencies ($f_{res}$) vs. the external magnetic field via the Kittel formula[29]:

$$f_{res} = (\frac{\gamma}{2\pi})[H_{res}(H_{res} + 4\pi M_{eff})]^{1/2} \quad (4)$$

in which $H_{res}$ is the magnetic field at the resonance condition, and $M_{eff}$ is the effective magnetization which contains the saturation magnetization and other anisotropy contributions. As shown in Figures 5a and 5b, the $4\pi * M_{eff}$ for 30 nm Py films with TaN capping layer are obtained to be ~10.4 and ~10.9 kG at 300 K and 5 K respectively. The temperature dependences of the $4\pi * M_{eff}$ for 3nm, 5 nm, and 30 nm Py films are shown in Figures 6a-6c. Around ~50 K, an anomaly in the effective magnetization for thin Py films (3 and 5 nm) is observed. Since we do not expect any steep change in Py's saturation magnetization at this temperature, the anomaly in $4\pi * M_{eff}$ should be caused by an anisotropy change which could be related to a spin reorientation. However, to fully understand the underlying mechanisms of the peak of the surface damping at ~ 50 K, further theoretical and experimental studies are needed.

## Conclusion

In summary, the thickness and temperature dependences of the Gilbert damping in Py thin films are investigated, from which the contribution due to the bulk damping and surface damping are clearly identified. Of particular interest, the bulk damping decreases monotonically as the temperature decreases, while the surface damping develops an enhancement peak at ~ 50 K, which could be related to a thermally induced spin reorientation for the surface magnetization of the Py thin films. This model is also consistent with the observation of an enhancement of the effective magnetization below ~50 K. Our experimental results will contribute to the understanding of the intrinsic and extrinsic mechanisms of the Gilbert damping in FM thin films.

## Methods

**Materials growth.** The Py thin films are deposited on ~25 nm $SiO_2$/Si substrates at room temperature in $3\times10^{-3}$ Torr argon in a magnetron sputtering system with a base pressure of ~ $1\times10^{-8}$ Torr. The growth rate of the Py is ~ 1 Å/s. To prevent *ex situ* oxidation of the Py film during the measurement, a ~ 20 Å TaN or $Al_2O_3$ capping layer is grown *in situ* environment. The TaN layer is grown by reactive sputtering of a Ta target in an argon-nitrogen gas mixture (ratio: 90/10). For $Al_2O_3$ capping layer, a thin Al (3 Å) layer is deposited first, and the $Al_2O_3$ is deposited by reactive sputtering of an Al target in an argon-oxygen gas mixture (ratio: 93/7).

**FMR measurement.** The FMR is measured using the vector network analyzer (VNA, Agilent E5071C) connected with a coplanar wave guide[30] in the variable temperature insert of a Quantum Design Physical Properties Measurement System (PPMS) in the temperature range from 300 to 2 K. The Py sample is cut to be 1 × 0.4 cm and attached to the coplanar wave guide with insulating silicon paste. For each temperature from 300 K to 2 K, the forward complex transmission coefficients ($S_{21}$) for the frequencies between 1 - 15 GHz are recorded as a function of the magnetic field sweeping from ~2500 Oe to 0 Oe.

**Contributions**

J.S. and W.H. proposed and supervised the studies. Y.Z. and Q.S. performed the FMR measurement and analyzed the data. T.S. and W.Y. helped the measurement. S.H.Y. and S.S.P.P. grew the films. Y.Z., J.S. and W.H. wrote the manuscript. All authors commented on the manuscript and contributed to its final version.

**Acknowledgements**

We acknowledge the fruitful discussions with Ryuichi Shindou, Ke Xia, Ziqiang Qiu, Qian Niu, Xincheng Xie and Ji Feng and the support of National Basic Research Programs of China (973 Grants 2013CB921903, 2014CB920902 and 2015CB921104). Wei Han also acknowledges the support by the 1000 Talents Program for Young Scientists of China.

**Competing financial interests**

The authors declare no competing financial interests.

## References:

1 Landau, L. & Lifshitz, E. On the theory of the dispersion of magnetic permeability in ferromagnetic bodies. *Phys. Z. Sowjetunion* **8**, 153 (1935).

2 Gilbert, T. L. A phenomenological theory of damping in ferromagnetic materials. *Magnetics, IEEE Transactions on* **40**, 3443-3449, doi:10.1109/TMAG.2004.836740 (2004).

3 Scheck, C., Cheng, L., Barsukov, I., Frait, Z. & Bailey, W. E. Low Relaxation Rate in Epitaxial Vanadium-Doped Ultrathin Iron Films. *Phys. Rev. Lett.* **98**, 117601 (2007).
4 Woltersdorf, G., Kiessling, M., Meyer, G., Thiele, J. U. & Back, C. H. Damping by Slow Relaxing Rare Earth Impurities in $Ni_{80}Fe_{20}$ *Phys. Rev. Lett.* **102**, 257602 (2009).
5 Radu, I., Woltersdorf, G., Kiessling, M., Melnikov, A., Bovensiepen, U., Thiele, J. U. & Back, C. H. Laser-Induced Magnetization Dynamics of Lanthanide-Doped Permalloy Thin Films. *Phys. Rev. Lett.* **102**, 117201 (2009).
6 Yin, Y., Pan, F., Ahlberg, M., Ranjbar, M., Dürrenfeld, P., Houshang, A., Haidar, M., Bergqvist, L., Zhai, Y., Dumas, R. K., Delin, A. & Åkerman, J. Tunable permalloy-based films for magnonic devices. *Phys. Rev. B* **92**, 024427 (2015).
7 Rantschler, J. O., McMichael, R. D., Castillo, A., Shapiro, A. J., Egelhoff, W. F., Maranville, B. B., Pulugurtha, D., Chen, A. P. & Connors, L. M. Effect of 3d, 4d, and 5d transition metal doping on damping in permalloy thin films. *J. Appl. Phys.* **101**, 033911 (2007).
8 Ingvarsson, S., Ritchie, L., Liu, X. Y., Xiao, G., Slonczewski, J. C., Trouilloud, P. L. & Koch, R. H. Role of electron scattering in the magnetization relaxation of thin $Ni_{81}Fe_{19}$ films. *Phys. Rev. B* **66**, 214416 (2002).
9 Hickey, M. C. & Moodera, J. S. Origin of Intrinsic Gilbert Damping. *Phys. Rev. Lett.* **102**, 137601 (2009).
10 Zhang, S. & Li, Z. Roles of Nonequilibrium Conduction Electrons on the Magnetization Dynamics of Ferromagnets. *Phys. Rev. Lett.* **93**, 127204 (2004).
11 Kuneš, J. & Kamberský, V. First-principles investigation of the damping of fast magnetization precession in ferromagnetic 3d metals. *Phys. Rev. B* **65**, 212411 (2002).
12 Kamberský, V. Spin-orbital Gilbert damping in common magnetic metals. *Phys. Rev. B* **76**, 134416 (2007).
13 Hankiewicz, E. M., Vignale, G. & Tserkovnyak, Y. Inhomogeneous Gilbert damping from impurities and electron-electron interactions. *Phys. Rev. B* **78**, 020404 (2008).
14 Brataas, A., Tserkovnyak, Y. & Bauer, G. E. W. Scattering Theory of Gilbert Damping. *Phys. Rev. Lett.* **101**, 037207 (2008).
15 Arias, R. & Mills, D. L. Extrinsic contributions to the ferromagnetic resonance response of ultrathin films. *Phys. Rev. B* **60**, 7395-7409 (1999).
16 Walowski, J., Müller, G., Djordjevic, M., Münzenberg, M., Kläui, M., Vaz, C. A. F. & Bland, J. A. C. Energy Equilibration Processes of Electrons, Magnons, and Phonons at the Femtosecond Time Scale. *Phys. Rev. Lett.* **101**, 237401 (2008).
17 Stiles, M. D. & Miltat, J. in *Spin Dynamics in Confined Magnetic Structures III* Vol. 101 *Topics in Applied Physics* (eds Burkard Hillebrands & André Thiaville) Ch. 7, 225-308 (Springer Berlin Heidelberg, 2006).
18 Celinski, Z., Urquhart, K. B. & Heinrich, B. Using ferromagnetic resonance to measure the magnetic moments of ultrathin films. *J. Magn. Magn. Mater.* **166**, 6-26 (1997).
19 Barati, E., Cinal, M., Edwards, D. M. & Umerski, A. Gilbert damping in magnetic layered systems. *Phys. Rev. B* **90**, 014420 (2014).
20 Tserkovnyak, Y., Brataas, A., Bauer, G. E. W. & Halperin, B. I. Nonlocal magnetization dynamics in ferromagnetic heterostructures. *Rev. Mod. Phys.* **77**, 1375-1421 (2005).
21 Liu, Y., Yuan, Z., Wesselink, R. J. H., Starikov, A. A. & Kelly, P. J. Interface Enhancement of Gilbert Damping from First Principles. *Phys. Rev. Lett.* **113**, 207202 (2014).

22 Starikov, A. A., Kelly, P. J., Brataas, A., Tserkovnyak, Y. & Bauer, G. E. W. Unified First-Principles Study of Gilbert Damping, Spin-Flip Diffusion, and Resistivity in Transition Metal Alloys. *Phys. Rev. Lett.* **105**, 236601 (2010).

23 Mankovsky, S., Ködderitzsch, D., Woltersdorf, G. & Ebert, H. First-principles calculation of the Gilbert damping parameter via the linear response formalism with application to magnetic transition metals and alloys. *Physical Review B* **87**, 014430 (2013).

24 Bailey, W., Kabos, P., Mancoff, F. & Russek, S. Control of magnetization dynamics in $Ni_{81}Fe_{19}$ thin films through the use of rare-earth dopants. *Magnetics, IEEE Transactions on* **37**, 1749-1754 (2001).

25 Rantschler, J. O., Maranville, B. B., Mallett, J. J., Chen, P., McMichael, R. D. & Egelhoff, W. F. Damping at normal metal/permalloy interfaces. *Magnetics, IEEE Transactions on* **41**, 3523-3525 (2005).

26 Luo, C., Feng, Z., Fu, Y., Zhang, W., Wong, P. K. J., Kou, Z. X., Zhai, Y., Ding, H. F., Farle, M., Du, J. & Zhai, H. R. Enhancement of magnetization damping coefficient of permalloy thin films with dilute Nd dopants. *Phys. Rev. B* **89**, 184412 (2014).

27 Ghosh, A., Sierra, J. F., Auffret, S., Ebels, U. & Bailey, W. E. Dependence of nonlocal Gilbert damping on the ferromagnetic layer type in ferromagnet/Cu/Pt heterostructures. *Appl. Phys. Lett.* **98**, 052508 (2011).

28 Sierra, J. F., Pryadun, V. V., Russek, S. E., García-Hernández, M., Mompean, F., Rozada, R., Chubykalo-Fesenko, O., Snoeck, E., Miao, G. X., Moodera, J. S. & Aliev, F. G. Interface and Temperature Dependent Magnetic Properties in Permalloy Thin Films and Tunnel Junction Structures. *Journal of Nanoscience and Nanotechnology* **11**, 7653-7664 (2011).

29 Kittel, C. On the Theory of Ferromagnetic Resonance Absorption. *Phys. Rev.* **73**, 155 (1948).

30 Kalarickal, S. S., Krivosik, P., Wu, M., Patton, C. E., Schneider, M. L., Kabos, P., Silva, T. J. & Nibarger, J. P. Ferromagnetic resonance linewidth in metallic thin films: Comparison of measurement methods. *J. Appl. Phys.* **99**, 093909 (2006).

## Figure Captions

**Figure 1. Measurement of Gilbert damping in Py thin films via ferromagnetic resonance (Py thickness = 30 nm). a**, Ferromagnetic resonance spectra of the absorption for 30 nm Py thin films with TaN capping layer at gigahertz frequencies of 4, 6, 8, 10 and 12 GHz at 300 K after normalization by background subtraction. **b, c,** The half linewidths as a function of the resonance frequencies at 300 K and 5 K respectively. The red solid lines indicate the fitted lines based on equation (3), where the Gilbert damping constants could be obtained.

**Figure 2. Temperature dependence of the Gilbert damping of Py thin films with TaN capping. a**, The temperature dependence of the Gilbert damping for 3, 5, 10, 15, 20, 30, and 50 nm Py films. **b,** The Gilbert damping as a function of the Py thickness, $d$, measured at 300 K. **c,** The Gilbert damping as a function of $1/d$ measured at 300 K. The linear fitting corresponds to equation (4), in which the slope and the intercept are related to the surface contribution and bulk contribution to the total Gilbert damping. Error bars correspond to one standard deviation.

**Figure 3. Bulk and surface damping of Py thin films with TaN capping layer. a, b,** The temperature dependence of the bulk damping and surface damping, respectively. The inset table summarizes the experimental values reported in early studies. Error bars correspond to one standard deviation.

**Figure 4. Comparison of the Gilbert damping of Py films with different capping layers. a, b,** Temperature dependence of the Gilbert damping of Py thin films with TaN capping layer

(blue) and $Al_2O_3$ capping layer (green) for 5 nm Py and 30 nm Py, respectively. Error bars correspond to one standard deviation.

**Figure 5. Measurement of effective magnetization in Py thin films via ferromagnetic resonance (Py thickness = 30 nm). a, b,** The resonance frequencies vs. the resonance magnetic field at 300 K and 5 K, respectively. The fitted lines (red curves) are obtained using the Kittel formula.

**Figure 6. Effective magnetization of Py films as a function of the temperature. a, b, c,** Temperature dependence of the effective magnetization of Py thin films of a thickness of 3 nm, 5 nm and 30 nm Py respectively. In **b, c**, the blue/green symbols correspond to the Py with $TaN/Al_2O_3$ capping layer.

Figure 1

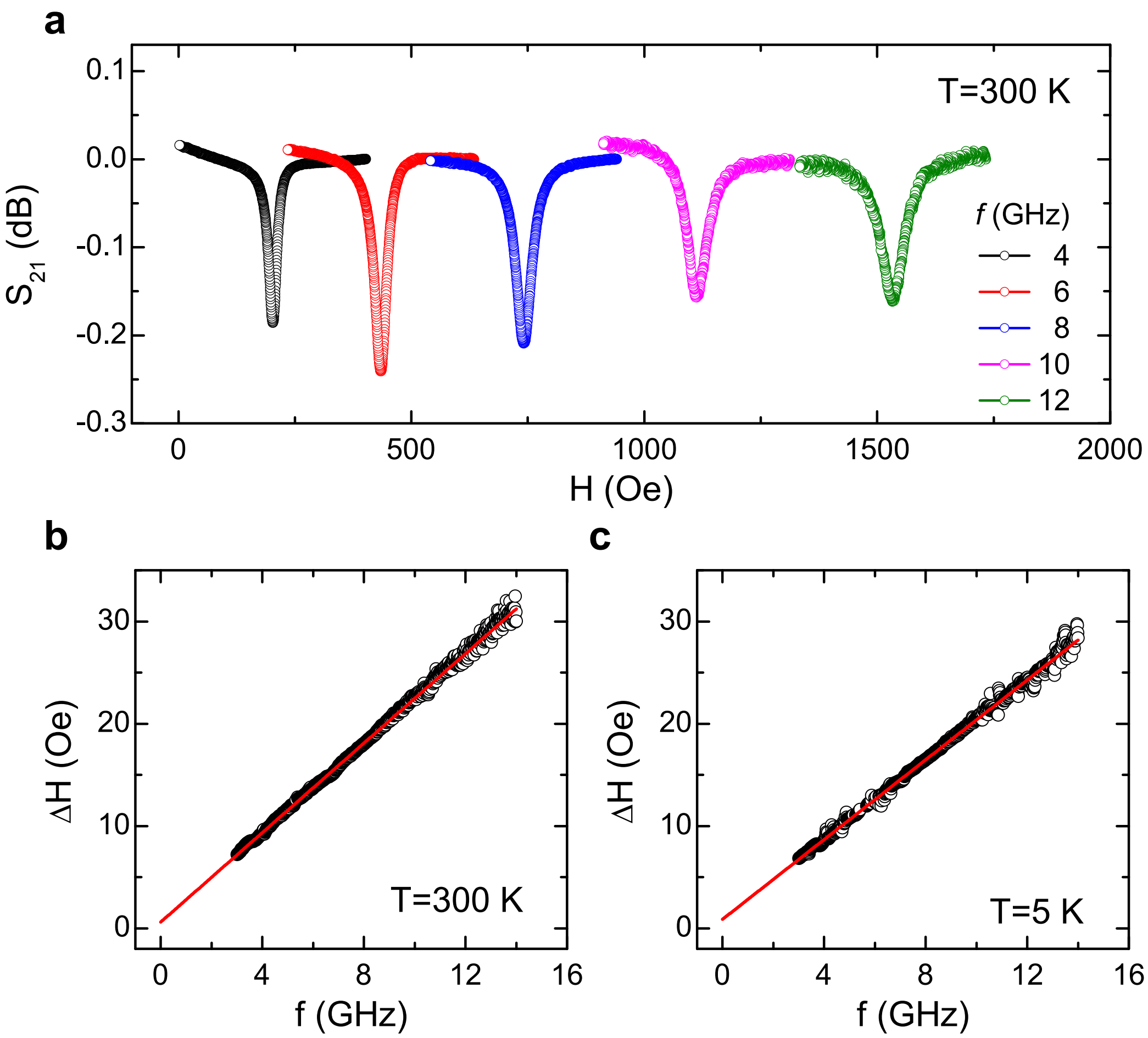

Figure 2

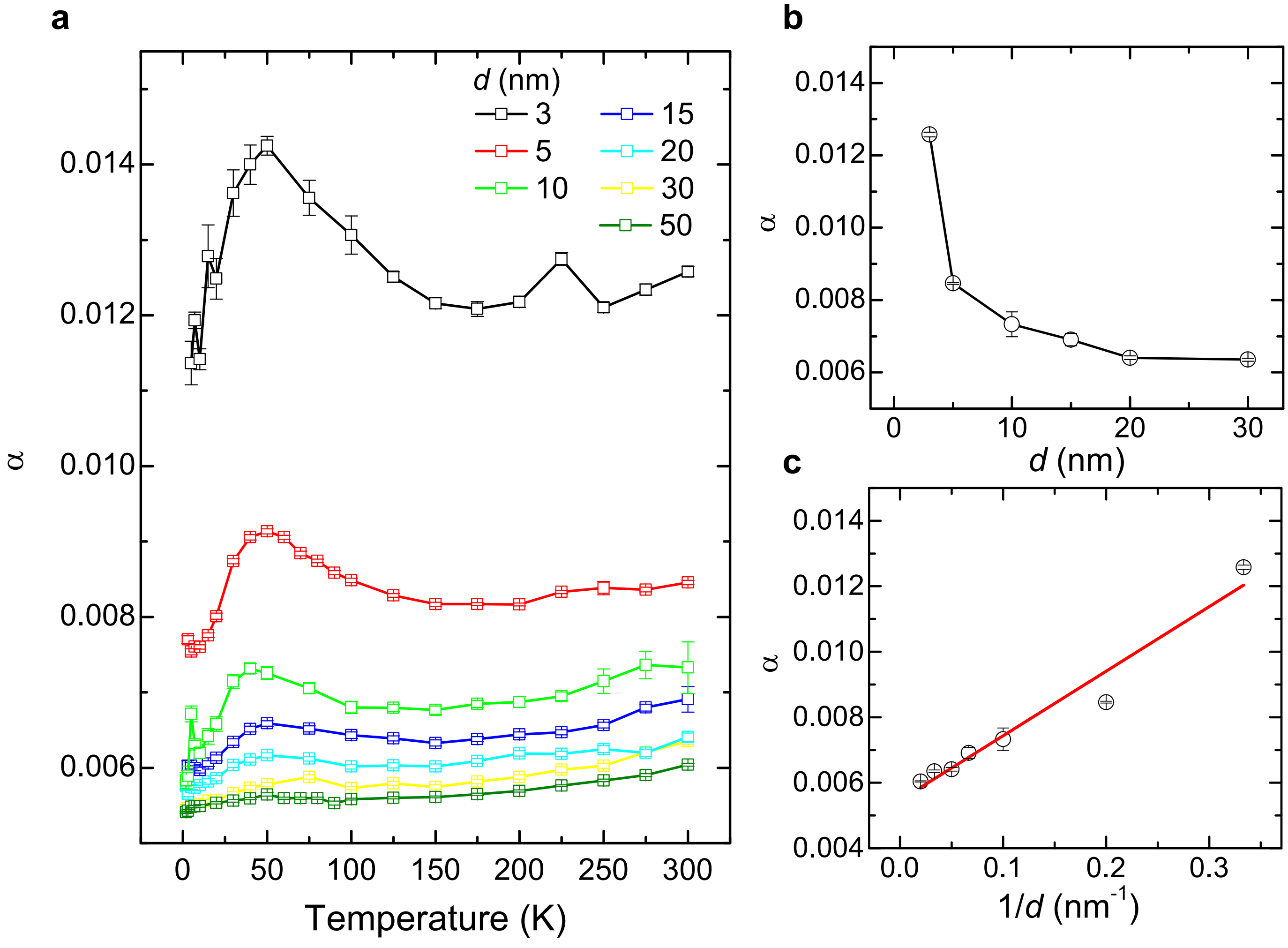

Figure 3

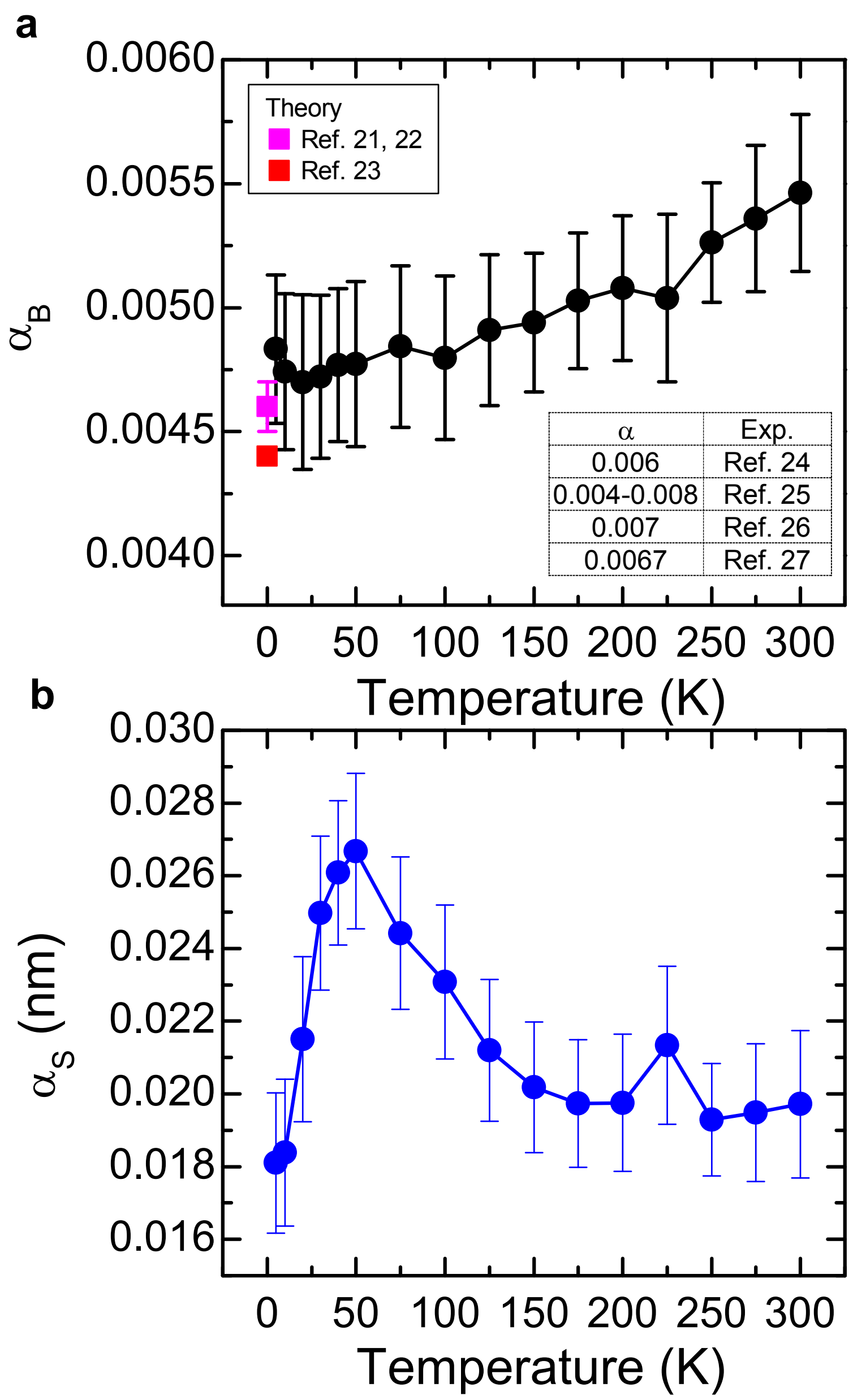
a
0.0060
0.0055
0.0050
0.0045
0.0040
$\alpha_B$
Theory
Ref. 21, 22
Ref. 23
α | Exp.
0.006 | Ref. 24
0.004-0.008 | Ref. 25
0.007 | Ref. 26
0.0067 | Ref. 27
0 50 100 150 200 250 300
Temperature (K)
b
0.030
0.028
0.026
0.024
0.022
0.020
0.018
0.016
$\alpha_S$ (nm)
0 50 100 150 200 250 300
Temperature (K)

Figure 4

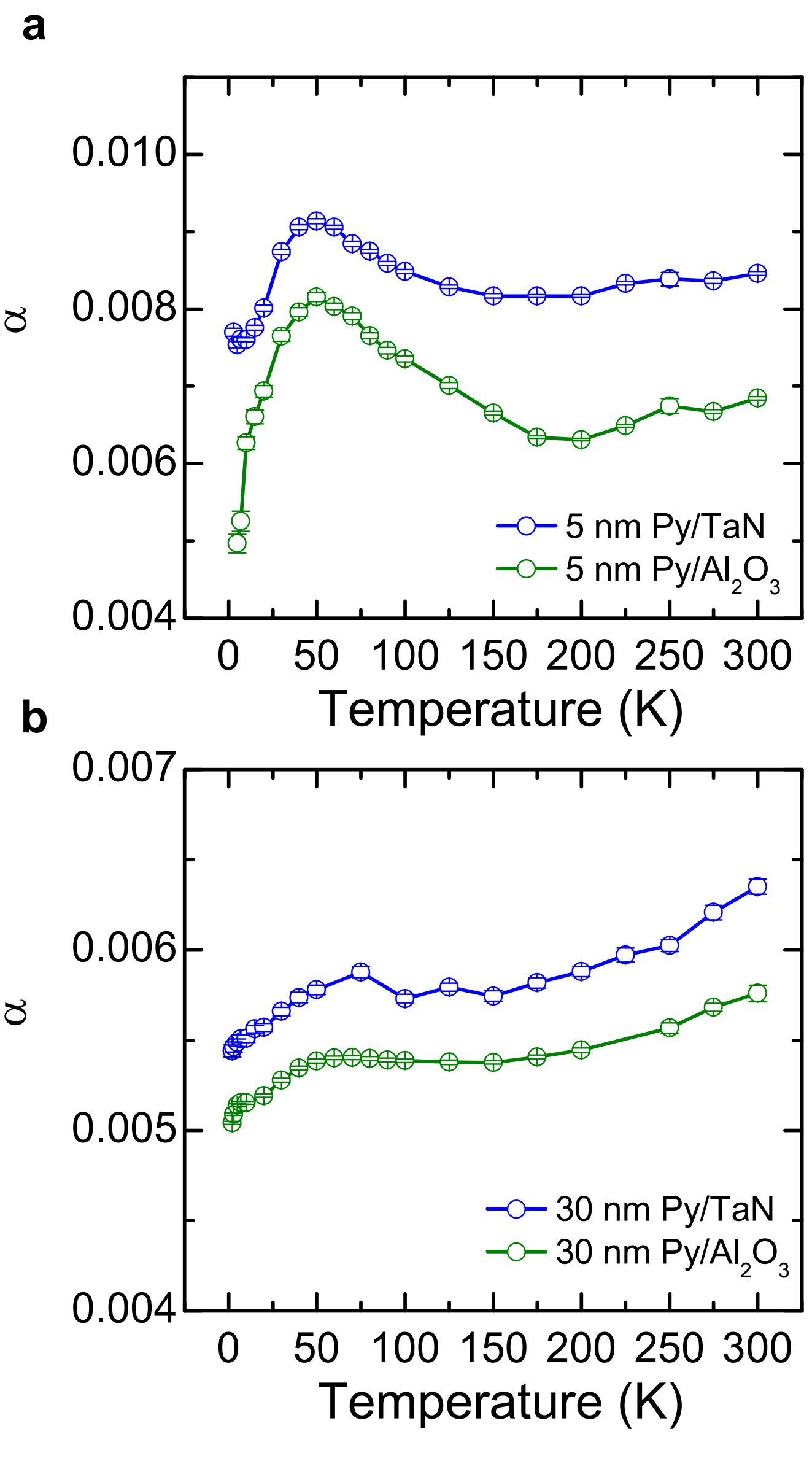

Figure 5

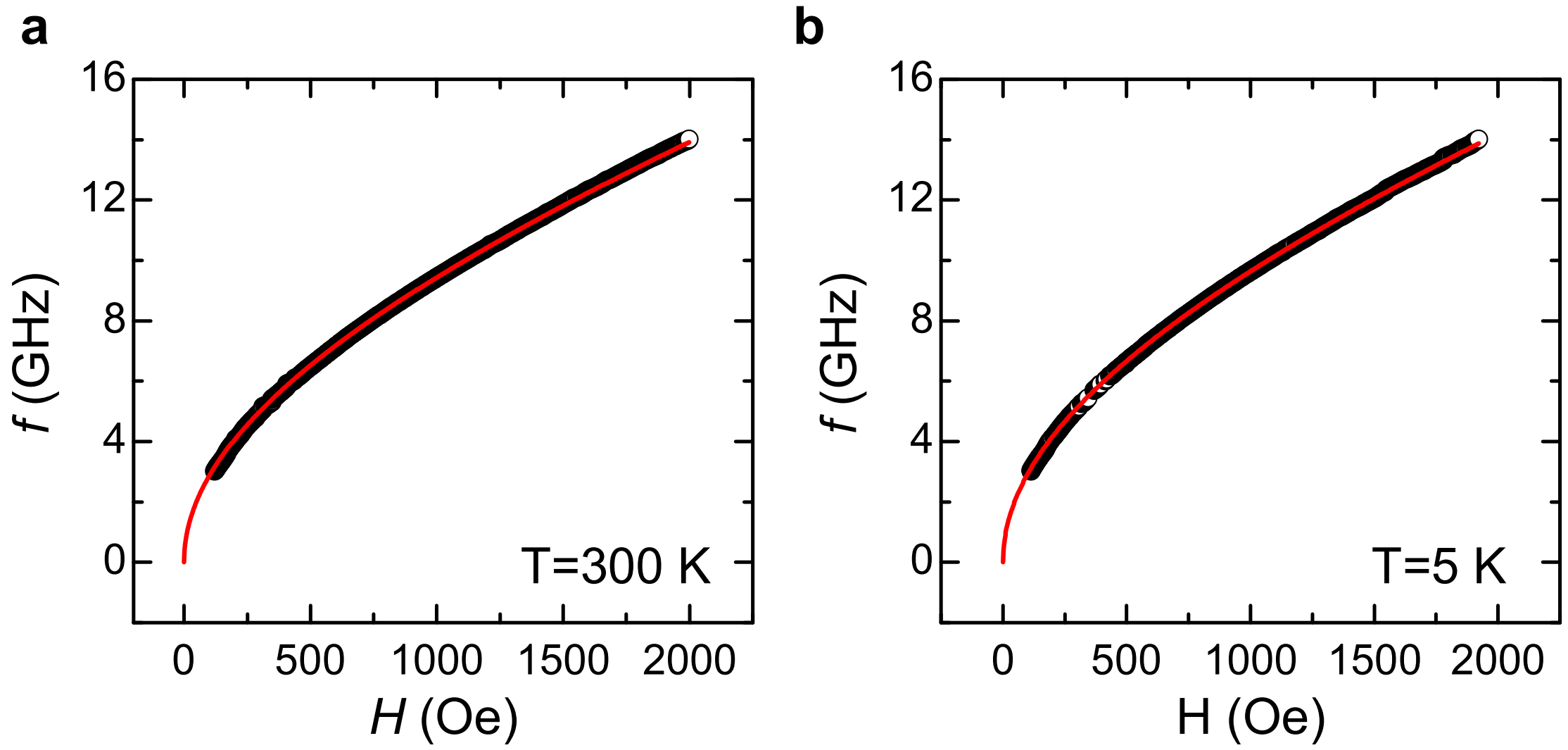

Figure 6

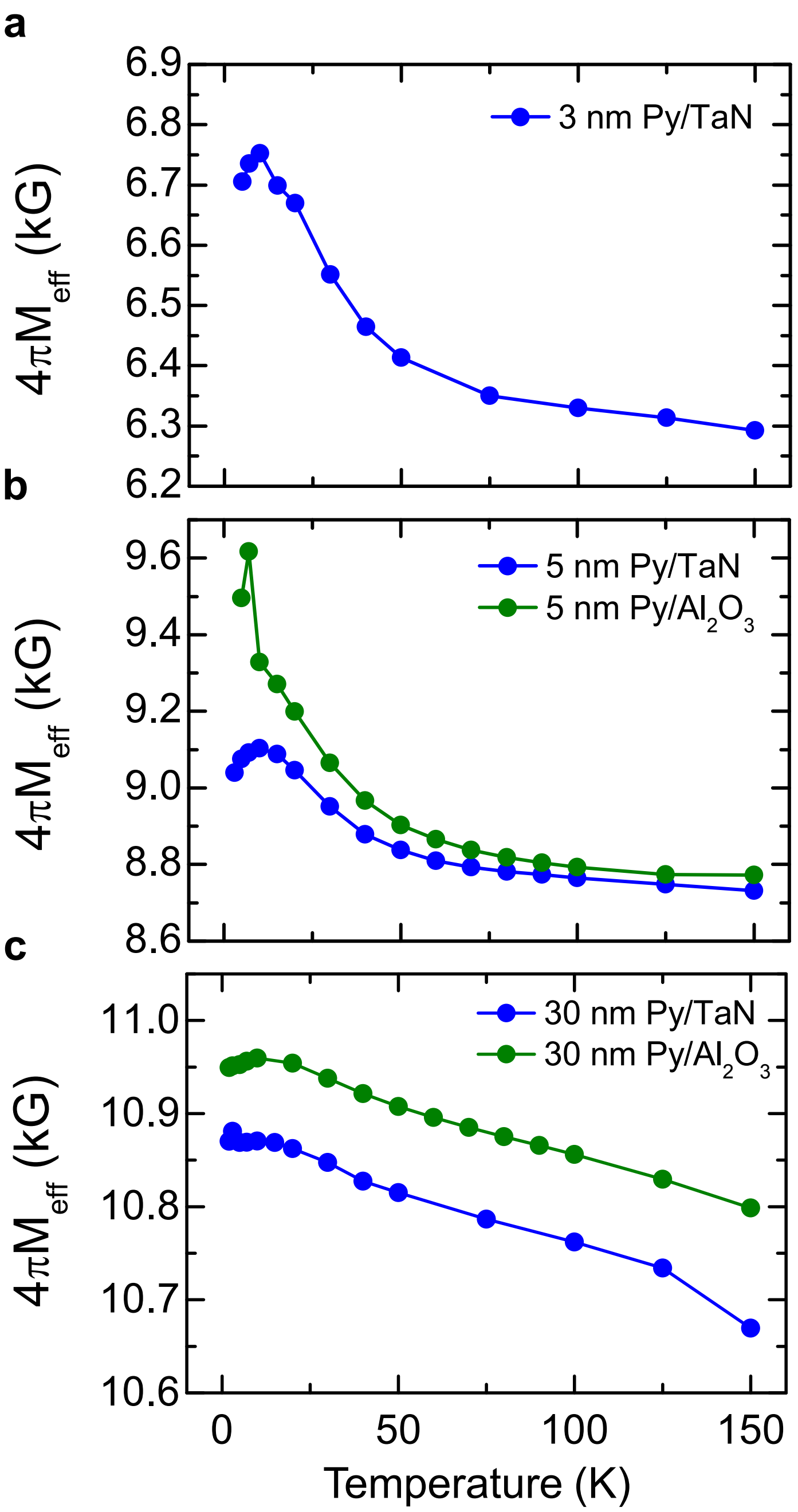